\documentstyle[12pt]{article}
\newcommand{\be}{\begin{equation}}
\newcommand{\ee}{\end{equation}}
\begin{document}
\date{\today}
\pagestyle{plain}
\title{Theory of the magnetronic laser}
\author{Miroslav Pardy\\
Institute of Plasma Physics ASCR\\
Prague Asterix Laser System, PALS\\
Za Slovankou 3, 182 21 Prague 8, Czech Republic\\
and\\
Department of Physical Electronics \\
Masaryk University \\
Kotl\'{a}\v{r}sk\'{a} 2, 611 37 Brno, Czech Republic\\
e-mail:pamir@physics.muni.cz}
\date{\today}
\maketitle
\vspace{50mm}

\begin{abstract}
We analyze the motion of an electron in the crossed electromagnetic field
of the planar magnetron. Numerically we calculate the parameters of
cycloid along which electron moves. We determine the total power of
radiation of an
electron moving in the magnetron and
the power spectrum generated by a single electron and by
the system of N electrons
moving coherently in the electromagnetic field of planar magnetron.
We argue that for large N, and high intensity of
electric and magnetic fields,
the power of radiation of such magnetronic laser, MAL,
can be sufficient for application in
the physical, chemical, biological and medicine sciences. In  medicine,
the magnetronic laser,
can be used for the therapy of the localized cancer tumors. The
application of the MAL in CERN as an ion source for LHC is not excluded.

\end{abstract}

\newpage

\baselineskip 15 pt

\section{Introduction}

Any of devices that produces an
intensive beam of light not only of a very pure single colour but all
colours is a laser.
While the gas lasers and masers produce practically a single colour, the
so called lasers on the free electrons such as undulators and wigglers
produce all spectrum of colours. These spectra are
generated in the form of so called synchrotron radiation because they
are generated during the
motion of electron when moving in the magnetic field.

The electrons moving inside the planar magnetron
evidently produce also the synchrotron
radiation because planar magnetron is a system of two metal charged plates
immersed into homogenous magnetic field which is perpendicular to the
homogenous electric field between plates.
The planar magnetron producing photons
has the same function as the free electron laser such as wiggler or undulator.
So, we denote such device by word magnetronic laser, or shortly MAL.
The motion of electrons
is determined by the equation of motion following from
electrodynamic of charges in homogenous electric and magnetic field.
The produced energy of electrons is given by the Larmor formula.

History teaches us that Van de Graaff (1931), Cockcroft and Walton
(1932) used the high voltage principle system
for the acceleration of particles and that the next revolution in
acceleration physics was
caused by Lawrence (Lawrence et al., 1932) who replaced the high
voltage acceleration
by repeated acceleration of particles moving along the circle. Here,
we use the high voltage
idea with the difference, to obtain synchrotron radiation only
and not to obtain high energy particles.

Although the radiation of one electron from one trajectory is very weak,
the total radiation of a MAL is not weak
because the process of radiation can be
realized by many electrons moving along the same trajectory.
There is no problem for today technology
to construct MAL of high intensity electric and magnetic fields.
We hope that the radiation output
of such device is sufficient large for application in an
physical laboratories.

We argue that MAL can be applied for instance in the therapy of cancer and
it means that in the specific situations it can be used instead of Leksell
gamma knife. On the other hand, the application of MAL in CERN as
the ion source for LHC is not excluded.

\section{The nonrelativistic motion of an electron
in the  field configuration of the planar magnetron}

We shall identify the direction of
magnetic field ${\bf H}$ with  $z$-axis and he direction of
electric field  ${\bf E}$ is along the
$y$-axis. Then, the
nonrelativistic equation of motion of an electron is as follows:

$$m\dot{\bf v} = e{\bf E} + \frac {e}{c}({\bf v}\times {\bf H}).\eqno(1)$$

This equation can be rewritten in the separate coordinates as follows
(we do not write the $z$ coordinate):

$$m\ddot x = \frac {e}{c} \dot y H,\eqno(2)$$

$$m\ddot y = eE_{y} - \frac {e}{c}\dot x H.\eqno(3)$$

Multiplying the equation (3) by $i$ and combining with the
equation (2), we find

$$\frac {d}{dt}(\dot x + i\dot y) + i\omega_{0}(\dot x + i\dot y)  =
i\frac {e}{m}E_{y},\eqno(4)$$
where

$$\omega_{0} = eH/mc.\eqno(5)$$

Function $\dot x + i\dot y$ can be considered as
the unknown, and as a such is equal to the sum of the integral of the same
differential equation without the right-hand term and a particular integral
of the equation with the right-hand term. The first integral is
$ae^{-i\omega_{0}t}$,
and the second integral is a constant which can be eliminated
from eq. (4) as $eE_{y}/m\omega_{0} = cE_{y}/H$. Thus

$$\dot x + i\dot y = ae^{-i\omega_{0} t} + \frac {cE_{y}}{H}.\eqno(6)$$

The constant $a$ is in general complex. So, we can write it in the form
$a = be^{i\alpha}$ where $b$ and $\alpha$ are real constants. We see that
since $a$ is multiplied by $e^{-i\omega_{0} t}$, we can, by a suitable choice
of the time origin, give the phase $\alpha$ any arbitrary value. We choose $a$
to be real. Then, breaking up $\dot x + i\dot y$ into real and imaginary
parts, we find

$$\dot x = a \cos\omega_{0} t + \frac {cE_{y}}{H}, \quad
\dot y = -a\sin\omega_{0} t.\eqno(7)$$

At $t = 0$ the velocity is along the x-axis (for $a \ne -cE_{y}/H$ ).

As we have said, all the formulas of this section assume that the velocity of
the particle is small compared with the velocity of light.
If we calculate the average value of $x$ and $y$, we get

$$<{\dot x}> = \frac {cE_{y}}{H},\quad  <{\dot y}> = 0, \eqno(8)$$
and, if we we define the drift velocity by relation

$$v_{drift} = \frac {cE_{y}}{H}, \eqno(9)$$
then, from the nonrelativistic condition $v_{drift} \ll 1$, we see that for
this to be so, it is necessary in particular that the electric and magnetic
intensities satisfy the following condition:

$$\frac {E_{y}}{H} \ll  1,\eqno(10)$$
while the absolute magnitudes of $E$ and $H$ can be arbitrary.

Integrating equation (7), and choosing the constant of integration so
that at $t = 0, x = y =0$, we obtain

$$x = \frac {a}{\omega_{0}} \sin\omega_{0} t + \frac {cE_{y}}{H}t, \quad
y = \frac {a}{\omega_{0}}(\cos\omega_{0} t - 1).\eqno(11)$$

These equations define in a parametric form a trochoid. Depending on whether
$a$ is larger or smaller in absolute value than the quantity
$cE_{y}/H$, the projection of the trajectory on the plane $xy$ have
different form.

If $a = -cE_{y}/H$, then, equation (7) becomes:

$$x = \frac {cE_{y}}{\omega_{0} H}(\omega_{0} t - \sin\omega_{0} t),\quad
y = \frac {cE_{y}}{\omega_{0} H}(1 - \cos \omega_{0} t). \eqno(12)$$

These equations are the parametric equation of cycloid in the plane $xy$.
The cycloid can be formed also mechanically as a curve traced
out by a point on the circumference of a circle that rolls without
slipping along a straight line. Constant $R = cE_{y}/\omega_{0} H$
is equal to the radius of the
rolling circle. The distance from the point with parameter $t = 0$ to the
point with parameter $t = 2\pi/\omega_{0}$ is $2\pi R$. The drift velocity
is $R \omega_{0}$ and it is the velocity of the center of the circle.

On the other hand we see that the average value of $\dot x$ is just the
drift velocity $v_{drift}$ in equation (9).

The generalization to the relativistic
situation can be performed using the approach of Landau et all. (1962).

The aim of this article is to investigate the radiation of electron
in case that the motion is just along cycloid, i. e.
the trajectory of an electron in the planar magnetron. We want to show
that such configuration of fields  is an
experimental device which can be used as the new effective source
of synchrotron radiation.

For the sake of simplicity we calculate spectrum of radiation of MAL in
the system $S'$ which moves with regard to the
magnetron at the drift velocity $v_{drift} = R\omega_{0}$.
Then, all electron trajectories are circles
and the radiation of this system is the synchrotron radiation.
Then, we can transform the power spectrum to the
system joined with the planar magnetron. However, we shall see that the
spectrum is modified slightly, because the drift velocity is not
relativistic in our situation.

The motion of electrons differs form the motion of electrons
in wigglers and undulators
where the trajectories of electrons are not cycloids and the motion is
highly relativistic.

It is also possible to realize the relativistic motion of electrons
in MAL. However, it needs special experimental conditions which are very expensive.

\section{Physical design of MAL}

The expression for the instantaneous power radiated by the nonrelativistic
charge $e$ undergoing an acceleration ${\bf a} = d{\bf v}/dt$ is given by the
Larmor formula as follows:

$$P = \frac {2}{3}\frac {e^{2}}{c^{3}}\left(\frac {d{\bf v}}{dt}\right)^{2}
 = \frac {2}{3}\frac {e^{2}}{m^{2}c^{3}}\left(\frac {d{\bf p}}{dt}\right)^{2}.
\eqno(13)$$

So we see that only electrons with undergoing big acceleration  can produce
sufficient power of radiation of photons. If we consider electrons in the
system $S'$ with the drift velocity $v = v_{drift} = cE_{y}/H$, then the
equations of the trajectory are as follows:

$$x = -\frac {cE_{y}}{\omega_{0}H}\sin\omega_{0}t, \quad
y = \frac {cE_{y}}{\omega_{0}H}(1 - \cos\omega_{0} t).\eqno(14)$$

They can be rewritten in the form:

$$x^{2} + \left(y - \frac {cE_{y}}{\omega_{0}H}\right)^{2} = \left(
\frac {cE_{y}}{\omega_{0}H}\right)^{2},\eqno(15)$$
which is evidently an equation of a circle with the radius

$$R = \frac {cE_{y}}{\omega_{0}H} = \left(\frac{eH}{mc}\right)^{-1}
\frac {cE_{y}}{H}
 = \frac {mc^{2}E_{y}}{eH^{2}}.\eqno(16)$$

The corresponding tangential velocity is

$$v_{t} = \sqrt{{\dot x}^{2} + {\dot y}^{2}} = \frac {cE_{y}}{H} = v_{drift}.
\eqno(17)$$

The corresponding acceleration is

$$ \sqrt{{\ddot x}^{2} + {\ddot y}^{2}} = \frac {eE_{y}}{m},
\eqno(18)$$
which corresponds to the Newton law, mass $\times$ acceleration  = force =
$eE_{y}$, which agrees with the introducing the electrostatic intensity $E_{y}$
inside the planar magnetron.

Big acceleration gives big radiation and also big tangential velocity
for the given intensity of the magnetic field. However,
big intensity of the magnetic field reduces the big
circle to the small size.

If we describe the motion of electrons in planar magnetron in the SI system
of units (Jackson, 1998) and in the system $S'$,
then, we write the following equations of motion:

$$x = -\frac {E_{y}}{\omega_{0}B}\sin\omega_{0}t, \quad
y = \frac {E_{y}}{\omega_{0}B}(1 - \cos\omega_{0} t),
\eqno(19)$$
where $B$ is the magnetic induction and

$$\omega_{0}  = \frac {eB}{m}.\eqno(20)$$

Then, for the radius of a circle, we get

$$R = \frac {E_{y}}{\omega_{0}B} = \left(\frac{eB}{m}\right)^{-1}
\frac {E_{y}}{B}
 = \frac {mE_{y}}{eB^{2}}.\eqno(21)$$

The corresponding tangential velocity is

$$v_{t} = \sqrt{{\dot x}^{2} + {\dot y}^{2}} = \frac {E_{y}}{B} = v_{drift}.
\eqno(22)$$

Let us determine acceleration of electron with the following parameters:
$E_{y} = 10^{6}\; {\rm V}\;{\rm m^{-1}}$, with $e = 1.6\times 10^{-19}{\rm C}$,
and $m = m_{e} = 9.1\times 10^{-31} {\rm kg}$. We get

$$\frac {eE_{y}}{m} = \frac {1.6 \times 10^{-19} \times 10^{6}}
{9.1 \times 10^{-31}} = \frac {1.6}{9.1} \times 10^{18} \approx
1.75 \times 10^{17} {\rm m}\; {\rm s^{-2}}.\eqno(23)$$

In the SI system the radiation power is as follows (Wille, 2000):

$$P = \frac {e^{2}}{6\pi \varepsilon_{0}c^{3}}\times
\left(\frac {eE_{y}}{m}\right)^{2},\eqno(24)$$
where

$$\varepsilon_{0} = 8.8\times 10^{-12} {\rm A}\; {\rm s}\;
{\rm V^{-1}}\; {\rm m}^{-1}.\eqno(25)$$

For $E_{y} = 10^{6}\;{\rm V}\; {\rm m^{-1}}$, with
$e = 1.6\times 10^{-19}\;{\rm C}$,
$c = 3\times 10^{8}\;{\rm m}\; {\rm s^{-1}}$ and $m = m_{e} =
 9.1\times 10^{-31}\;{\rm kg}$, we get
using the value of acceleration determined in equation (23):

$$P = \frac {(1.6)^{2}\times (1.75)^{2}}{6\times 3.14 \times 8.8
\times 27}\times (10^{-19 + 17})^{2} \times 10^{12 - 24}\;
{\rm J} \; {\rm s^{-1}} \approx 1.75 \times 10^{-20}\; {\rm W}.
\eqno(26)$$

Of course this is very small radiation, however this is only for one electron,
or, in other words for one elementary charge. If we prepare regime with many
charges, say $q = 10^{10}e$, then the situation will be substantially
different. We obtain $P = 1.75 \;{\rm W}$.

We can also easily determine the radius of circle in the system $S'$. The
corresponding formula in the SI system of units is given by equation (21): $R =
mE_{y}/eB^{2}$, where $B$ is the magnetic induction. For $B$ = 1T, we get:

$$R = \frac {9.1}{1.6}\times 10^{-31+6+19} \approx 5.7 \times
10^{-6}\; {\rm m}.\eqno(27)$$

It follows from the last formula, that the radius of circle is very
small and it means that the external observer sees the radiating
``straight line'',
which is parallel with the $x$-axis.
Every point of this ``straight line'' radiates, however the
enhancement of radiation is only at the direction of ``straight line''.

If we are interested also in the value of the drift velocity for our
parameters of the MAL, then we get:

$$v_{drift} = \frac {E}{B} = 10^{6}\; {\rm m\;s^{-1}}.\eqno(28)$$

So we see, that we have chosen the parameters in a such way that
the drift velocity is substantially smaller than the velocity of light.

The quantity $\omega_{0}$ in the considered situation with $B$ = 1T is
determined as follows:

$$\omega_{0} = \frac{eB}{m} = \frac{1.6}{9.1}\times 10^{12}\;
{\rm s^{-1}}\approx  1.7 \times 10^{11} \; {\rm s^{-1}}.\eqno(29)$$

For current which is formed by charges moving along the trajectory,
we can derive simple formula using the physical ingrediences in the preceding
text.

$$J = e(N/L)v_{drift} = e(1/2\pi R)v_{drift} = 1.6 \times 2.8\times
10^{-9}{\rm  A} \approx
4.5 \times 10^{-9}\;{\rm  A}.\eqno(30)$$

We see, that the current generated by the high-intensity field with only one
electron at the arc of the cycloid is very small.
However for $e \rightarrow 10^{10}e$, we obtain $J = 45 \;{\rm A}$

If the distribution of the synchrotron is $P(\lambda)$, then, the
maximal intensity of radiation is is for the following
$\lambda = \lambda_{max}$ (Schwinger, 1949):

$$\lambda_{max} \approx (4\pi/3) R\left(\frac {mc^{2}}{W}\right)^{3},
\eqno(31)$$
where $R$ is the radius of the  circle and $W$ is the energy
of electron with the rest mass $m$ which moves along the circle.

If we assume that MAL works at above conditions then,
with the electron rest energy $mc^{2} = 0.5$ MeV, and $W \approx 0.5$ MeV, we get:

$$\lambda_{max}\approx  (4\pi/3) R \approx 23.9 \times 10^{-6}\;
{\rm m} = 23.9 \; \mu{\rm m}, \quad B = 1{\rm T},\eqno(32)$$
which is the infrared wavelength. For $B$ = 5T, we obtain
$$\lambda_{max} \approx (4\pi/3) R \approx 960 \; {\rm nm},\quad  B =  5{\rm T}. \eqno(33)$$

However, the synchrotron radiation is
generated in the form of the all wavelength and it means also for $\lambda <
\lambda_{max}$. It means that the planar magnetron generates also the
Roentgen radiation. Of course, the intensity of the Roentgen radiation
of one electron is very weak. However, in case with many
electrons, it can be very strong and it differs from the laser radiation,
which is produced in the optical frequencies. So, the practical application
of the planar magnetron as a source of the Roentgen, or, synchrotron radiation
is possible.

The planar magnetron is composed from the cathode and anode. If the cathode
is cold, then after application the voltage the emission of electrons occurs
accidentally from the arbitrary points of the cathode.
In order to establish only one starting point
of the emission of electrons, we connect  cathode with  the prismatical,
or conical
protrusion with the vertex between anode and cathode. Then, the electrons are
sucked from the vertex to the anode and move along the trajectory of
the constant geometrical form. Then, the planar magnetron as a source of
synchrotron radiation works.

In case of the thermal cathode, the initial velocities of electrons are determined by
the statistical law. The consequence of it is, that many different cycloids are
generated and the coherence of motion is broken.

\section{The spectral function of the radiation of relativistic electrons}

Hitherto, we considered the nonrelativistic electrons. Now, we
shall work with the relativistic electrons for which it is elaborated
the Schwinger formalism (source theory) which enables to perform calculations in a
simple  and brilliant form.
We shall derive the power spectrum formula of the synchrotron radiation
generated by the motion of an electron moving  in a planar magnetron.
We follow  Schwinger et al. (1976), but we will follow also the
author articles (Pardy, 1994, 1997).

Let us remark, that source theory methods
(Schwinger, 1970, 1973; Dittrich, 1978)  was initially constructed
for a description of high-energy particle physics experiments.
It was found that the
original formulation simplifies the calculations in
electrodynamics and gravity, where the interactions are mediated by
the photon and graviton, respectively. It simplifies particularly the
calculations with radiative corrections (Dittrich, 1978; Pardy, 1994) .

The basic formula of the Schwinger source theory is the so called
vacuum to vacuum amplitude:

$$\langle 0_{+}|0_{-} \rangle = e^{\frac{i}{\hbar}\*W},
\eqno(34)$$
where, for the case of the electromagnetic field in the medium, the action
$W$ is given by

$$W = \frac{1}{2c^2}\*\int\,(dx)(dx')J^{\mu}(x){D}_{+\mu\nu}(x-x')J^{\nu}(x'),
\eqno(35)$$
where

$${D}_{+}^{\mu\nu} = \frac{\mu}{c}[g^{\mu\nu} +
(1-n^{-2})\beta^{\mu}\beta^{\nu}]\*{D}_{+}(x-x'),
\eqno(36)$$
and $\beta^{\mu}\, \equiv \, (1,{\bf 0})$, $J^{\mu}\, \equiv \,(c\varrho,{\bf
J})$ is the conserved current, $\mu$ is the magnetic permeability of
the medium, $\epsilon$ is the dielectric constant of the medium, and
$n=\sqrt{\epsilon\mu}$ is the index of refraction of the medium.
Function ${D}_{+}$ is defined as follows (Schwinger et al., 1976):

$$D_{+}(x-x') =\frac {i}{4\pi^2\*c}\*\int_{0}^{\infty}d\omega
\frac {\sin\frac{n\omega}{c}|{\bf x}-{\bf x}'|}{|{\bf x} - {\bf x}'|}\*
e^{-i\omega|t-t'|}.
\eqno(37)$$

The probability of the persistence of vacuum follows from the vacuum
amplitude (34) in the following form:

$$|\langle 0_{+}|0_{-} \rangle|^2 = e^{-\frac{2}{\hbar}\*\rm Im\*W},
\eqno(38)$$
where ${\rm Im}\;W$ is the basis for the definition of the spectral
function $P(\omega,t)$ as follows:

$$-\frac{2}{\hbar}\*{\rm Im}\;W \;\stackrel{d}{=} \; -\,
\int\,dtd\omega\frac{P(\omega,t)}
{\hbar\omega}.
\eqno(39)$$

Now, if we insert equation (35) into eq. (39), we get,
after extracting $P(\omega,t)$, the following general expression
for this spectral function:

$$P(\omega,t) = -\frac{\omega}{4\pi^2}\*\frac{\mu}{n^2}\*\int\,d{\bf x}
d{\bf x}'dt'\left[\frac{\sin\frac{n\omega}{c}|{\bf x} -
{\bf x}'|}{|{\bf x} - {\bf x}'|}\right]\;\times $$

$$\cos[\omega\*(t-t')]\*[\varrho({\bf x},t)\varrho({\bf x}',t')
- \frac{n^2}{c^2}\*{\bf J}({\bf x},t)\cdot{\bf J}({\bf x}',t')].
\eqno(40)$$

Formula (40) was  obtained also by Schwinger from the classical
electrodynamics without using
source theory methods (Schwinger, 1945, 1949).

\section{Radiation of N electrons moving coherently in a planar magnetron}

We determine the power spectral formula in the system  $S'$
moving with the drift velocity $v_{drift} =
cE_{y}/H$ with regard to magnetron. In this case
motion of electron is not cycloid but the circle and we can repeat some
formulae which were used in the previous work of author (Pardy, 2000; 2002).
Such approach is meaningful because we can return to the system $S$ using the
formulas connecting frequencies and distribution in the
 system $S$ with frequencies
and photon distribution in the system $S'$ (Landau et al., 1962).

We will apply the formula (40) to the N-body system
with the same charged
particles which moves along the same circles with diameters
$R$ with the centres at points $0$, $2R$, $4R$, ...,  $2(N-1)R$ ,
where $N$ is the natural number.

The general approach involves the consideration of parameters of medium
and the possibility of involving the \v Cerenkov electromagnetic radiation
which is generated by a fast-moving charged particle
in a medium when its speed is faster than the speed of light in this medium.
It seems that in our case of MAL this radiation represents only the
academical interest, nevertheless the future  application cannot be excluded.
So, the connection with the experimental  observation of this radiation by
\v{C}erenkov (1934) and theoretical interpretation  by
Tamm and Frank (1937) in
the framework of classical electrodynamics is possible as
the generalization of theory of MAL. A source
theory description by Schwinger et al.
(1976) in the zero-temperature regime and the at the  finite-temperature
regime by Pardy (1989, 1995) can be also considered as meaningful.

So we write for the charge density $\varrho$ and for the current
density ${\bf J}$
of the N-body system:

$$\varrho({\bf x},t) = e\sum_{i=1}^{i=N}\*\delta\*({\bf x}-{\bf x_{i}}(t))
\eqno(41)$$
and

$${\bf J}({\bf x},t) = e\sum_{i=1}^{i=N}\*{\bf v}_{i}(t)
\*\delta\*({\bf x}-{\bf x_{i}}(t)),
\eqno(42)$$
where ${\bf x_{i}}(t)$ is the rajectory of electrons in MAL. In order to be
in the formal identity with the Schwinger approach, we perform elementary
transformation of variables in eq. (12) which has no influence on the spectrum
of emitted photons. In other words we make the following transitions:

$$ S \rightarrow S';\quad \omega_{0} \rightarrow -\omega_{0};\quad
y \rightarrow y - R; \quad \omega_{0}t \rightarrow \omega_{0}t +
\frac {\pi}{2}. \eqno(43)$$

Then, we can write ${\bf x_{i}}(t)$ in eqs. (41) and (42) in the form:

$${\bf x}_{i}(t) = i{\bf a} + {\bf x}(t) =  i{\bf a}
+ R({\bf i}\cos(\omega_{0}t) + {\bf j}\sin(\omega_{0}t), \quad i = 1, 2, 3,
..., N; \quad {\bf a} = (2R, 0, 0).\eqno(44)$$

From the physical situation follows with
($H = |{\bf H}|, W =$ energy of a particle)

$${\bf v}_{i} = d{\bf x}_{i}/dt =  d{\bf x}/dt =
{\bf v}(t), \hspace{5mm} \omega_{0} = v/R, \hspace{5mm}
R = \frac {\beta\*W}{eH}, \hspace{5mm}
\beta = v/c, \hspace{5mm} v = |{\bf v}|.
\eqno(45)$$

After insertion of eqs. (41) and (42) into eq. (40), and after some mathematical
operations we get

$$P(\omega',t) =
-\frac{\omega'}{4\pi^2}\*\frac{\mu}{n^2}e^{2}\*\int_{-\infty}^{\infty}\,
dt'\cos\omega' (t-t')
\left[1 - \frac {{\bf v}(t)\cdot {\bf v}(t')}{c^{2}}n^{2}\right]
\;\times $$

$$\sum_{i=1}^{N}\sum_{j=1}^{N}
\left\{\frac{\sin\frac {n\omega'}{c}|{\bf x}_{i}(t) -{\bf x}_{j}(t')|}
{|{\bf x}_{i}(t) -{\bf x}_{j}(t')|}\right\}.
\eqno(46)$$

Using $t' = t + \tau$, we get with
$ {\bf x}(t) = R({\bf i}\cos(\omega_{0}t) + {\bf j}\sin(\omega_{0}t)$:

$${\bf x}_{i}(t) -{\bf x}_{j}(t') =  (i-j){\bf a} + {\bf x}(t) -{\bf x}(t+\tau)
= (i-j){\bf a} + {\bf A},
\eqno(47)$$
where ${\bf A}\stackrel{d}{=} {\bf x}(t) -{\bf x}(t+\tau)$.

Using geometrical representation of vector ${\bf x}(t)$, we get:

$$|{\bf A}| = [R^{2} + R^{2} - 2RR\cos(\omega_{0}\tau)]^{1/2} =
2R\left|\sin\left(\frac {\omega_{0}\tau}{2}\right)\right|,
\eqno(48)$$
and

$$
{\bf v}(t)\cdot{}{\bf v}(t+\tau) = \omega^{2}_{0}R^{2}\cos\omega_{0}\tau.
\eqno(49)$$

The absolute values of the $x$ differences are as follows:

$$|{\bf x}_{i}(t) -{\bf x}_{j}(t')| = \left(
(i-j)^{2}a^{2} + 2(i-j){\bf a}\cdot {\bf A} + A^{2}\right)^{1/2}.\eqno(50)$$

Now, using the definition of ${\bf a}
\equiv  (2R, 0, 0)$ and $|{\bf A}|$, we get

$$|{\bf x}_{i}(t) -{\bf x}_{j}(t')| =
\left((i-j)^{2}4R^{2} + 8R^{2}(i-j)\cos\theta +
4R^{2}\sin^{2}\frac {\omega_{0}\tau}{2}
\right)^{1/2},\eqno(51)$$
where $\theta$ is an angle between ${\bf a}$ and ${\bf A}$.

Now, we are prepared to write the power spectrum formula for emission of
photons by the planar magnetron with N electrons. We have all ingredients
for determination of the final formula and we can follows the way of
author article (Pardy, 2002).
However, we see that the last formula is very complicated and it means that
the calculation of the power spectrum will be not easy. So, in such a
situation, we can use some approximation. One possible approximation is
to consider the contributions of terms with $i=j$. Then we get:

$$P_{N}\approx N\* P,\eqno(52)$$
where $P(\omega')$ was expressed by Schwinger et al. (1976) in the form

$$P(\omega') = \sum_{l=1}^{\infty}  \delta(\omega' - l\omega_{0})P_{l},
\eqno(53)$$
where

$$P_{l}(\omega',t) =
N \frac {e^2}{\pi\*n^2}\*\frac {\omega'\mu\omega_{0}}{v}\*
\left(2n^2\beta^2J'_{2l}(2ln\beta) -
(1 - n^2\*\beta^2)\*\int_{0}^{2ln\beta}dxJ_{2l}(x)\right),
\eqno(54)$$
where $N$ is the number of arcs forming the trajectory of electrons
with one electron at one arc.

Our aim is to apply the last formula to the situation of MAL,
where electrons move in a vacuum. In
this case we can put $\mu = 1, n = 1$. At the same time, with regard to the
equation (52), we get for N electrons:

$$P_{(N\; electrons)l} =
N \frac {e^2}{\pi}\*\frac {\omega'\omega_{0}}{v}\*
\left(2\beta^2J'_{2l}(2l\beta) -
(1 - \beta^2)\*\int_{0}^{2l\beta}dxJ_{2l}(x)\right).
\eqno(55)$$

If we take the idea that the discrete spectrum parametrized by number
$l$ is effectively continuous for $l\gg 1$, then, in such a case there is
an relation (Schwinger, 1949):

$$P(\omega') =
P_{(l = \omega'/\omega_{0})}\*\left(\frac {1}{\omega_{0}}\right).
\eqno(56)$$

Formulas (55), (56) concerns the situation with electrons
moving along the circular
trajectory. In other words, we work in the system which moves with regard to
the magnetron at the drift velocity $v_{drift}$. In the case that we are in the
system joined with the magnetron, all wave lengths are shifted to the
blue edge because of the Doppler effect. If we consider only
photons moving along the $x$-axis, then, we can use the formula for
the Doppler shift of  the following form (Rohlf, 1994):

$$\lambda = \lambda' \sqrt{\frac {1-\beta}{1+\beta }}.\eqno(57)$$
because we move toward the photon emission.

For the drift velocity $v_{drift} = E_{y}/B$, with the electric
field $10^{6}$ V/m and magnetic field 1T, we get
$\lambda/\lambda'\approx 1- \beta \approx 0.997,\; {\rm for}\;  \beta  =
10^{6}/(3\times 10^{8}) \approx 3.3 \times 10^{-3}\ll 1$. It represents
practically no shift, because of the small drift velocity.

The radiative corrections
obviously influence the spectrum (Schwinger, 1970 and
Pardy, 1994). Determination of this phenomenon is not the interest of
this article.

\section{The use of MAL at the therapy of cancer}

We know from the medicine literature (Concise Medical Dictionary,
2003), that
the cancer is any malignant tumor, including carcinoma and sarcoma. It arises
from the abnormal and uncontrolled division of cells that then invade and
destroy the surrounding tissues. Cancer tumor is composed from cancer cells
which we call metastases and which are spread by blood stream in the
lymphatic channels, thus setting up secondary tumors at sites distant from
the original tumor. There are many causative factors of cancer,
such as cigarette smoking, radiation, viruses and so on. In more than half of
all cancers a gene called p53 is deleted or impaired. Its normal function is
to prevent the uncontrolled division of cells. Treatment of cancer depends on
the type of tumor, the site of the primary tumor and the extend of spread.

We know, (Concise Medical Dictionary, 2003)
that the usual  methods of
the cancer therapy are chirurgical, biochemical and radiological.
Radiotherapy, or, therapeutic radiology is the treatment of desease with
penetrating radiation, such as X-rays, beta rays, gamma rays, proton beams,
pion beams and so on. Specially, high energy protons have ideal
characteristics for treating deep-seated tumors (Jones et al., 2001).
The rays are usually produced by machines, or by the
certain radioactive isotopes. Beams of radiation may be directed at a
deseased object from a distance of radioactive material. Well known
technique
is Leksell gamma knife (Leksell, 1951).  At present time it uses
aproximatelly 200 sources of the gamma rays where gamma rays are produced
by radionuclides $^{60}$Co. Many forms of cancer  are
destroyed by this type of radiotherapy.

At present time the successful method is to treat cancer
by the synchrotron radiation generated by synchrotrons,
betatrons and microtrons. We proposed in this article , the planar
magnetron  laser, MAL, as the new medicine element for treating cancer.
The size of the planar megnetron is in no case so large
as the synchrotron and it means the cancer tumors can be negated
ambulantly. So, the cancer therapy by MAL is hopeful.

\section{Discussion}

The generation of the synchrotron radiation by the planar magnetron,
forms the analogue of the wiggler and undulator generation of
radiation (Winick, 1987). However, while the wiggler and undulator
radiators needs the high-energy electrons from additional accelerator,
the planar magnetron works with electrons which are accelerated by own
magnetic and electric fields. The acceleration
depends on the intensity of electric field, and, the curvature of trajectory
depends also on  the magnetic field.
The spectrum of the magnetron radiation in system
$S'$  moving with the drift velocity $v_{drift} = cE_{y}/H$,
is the spectrum of the synchrotron radiation
of N electrons moving along the circles of the same radius. According to the
special theory of relativity, there is only the magnetic field in the system
$S'$. The MAL can work
with the single electron on the trajectory, or, with many electrons on
one trajectory or many trajectories. In case of many electrons
the radiation can be very intensive. The merit of the MAL
is in small size with regard to large synchrotrons.
On the other hand the opening angle of radiation is not so small
as in the  synchrotron because electrons in MAL are nonrelativistic.
The small opening angle is possible only for high energy electrons.
According to Winick (1987), if an electron is given a total
energy 5 GeV, the opening angle
over which synchrotron radiation is emitted is only 0.0001 radian, or about
0.006 degree. This can be regarded as a beam with the nearly parallel rays.
This is practically the same as the laser beam situation. The
wave length of the magnetronic photons is from zero to infinity.
If we want
to produce  maximal energy of photons at the very short length of
photons, then we must apply in MAL
very high magnetic field and very high voltage.
If we use high-frequency high-voltage alternate current generated by
Tesla transformer, then we must
also alternate
the magnetic field in order to keep the motion of electrons in one
direction. This can be easily realized, because the $x(t)$
is invariant with regard the simultaneous alternating the electric and
magnetic field, as we can see:

$$\frac{c E_{y}}{H} \rightarrow \frac{c(-E_{y})}{(-H)},
\quad \omega_{0} \rightarrow -\omega_{0}.\eqno(58)$$

The variable $y$ is not invariant with regard to the transformation (58). It is
$y(t) \rightarrow -y(t)$.
However, the prismatical protrusion causes that only trajectories
with the positive electric and magnetic fields are realized.

From formula (32) follows that MAL  produces all wave lengths
including synchrotron radiation and visible light.
Using optical filters we can leave only the short wave lengths.
Of course, then,  the total applied radiation is decreased.
In conclusion, we hope, that MAL will play fundamental role in all
branches of science.

\newpage

\begin{center}
{\bf References}
\end{center}
\vspace{10mm}
\noindent

Cockroft, J. D. and Walton, E. T. S. (1932). Experiments with high
velocity ions. {\it Proc. Royal Soc.} {\bf A 136}, 619 - 630 \\

{\it Concise Medical Dictionary}, (2003). Oxford university press.\\

\v{C}erenkov, P. A. (1934). The visible radiation of pure liquids caused by
$\gamma$-rays. {\it Comptes Rendus Hebdomaclaires des Scances de l'Academic
des Sciences} (USSR) {\bf 2}, 451. \\

Dittrich, W. (1978). Source methods in quantum field theory.
{\it Fortschritte der Physik} {\bf 26}, 289.\\

Jackson, J. D. (1998). {\it Classical Electrodynamics}, Third ed., John Wiley
\& Sons, New York, etc. \\

Jones, D. T. L. and Schreuder, A. N.  (2001). Magnetically scanned
proton therapy beams: rationales and principles.
{\it Radiation Physics and Chemistry} {\bf 61}, 615-618. \\

Lawrence, E. O., and Livingston, M. S. (1932). The production of high speed
light ions without use of high voltages. {\it Phys. Rev.} {\bf 40}, 19-35.
\\

Landau, L. D. and Lifshitz, E. M., (1962). {\it The Classical
Theory of Fields}, 2nd ed. Pergamon Press, Oxford. \\

Leksell, L. (1951). The stereotaxic method and radiosurgery of the brain.
{\it Acta chir. Scan.} {\bf 102}, 316-319.\\

Pardy, M. (1989). Finite-temperature \v{C}erenkov radiation.
{\it Phys. Lett. A} {\bf 134}(6), 357.\\

Pardy, M. (1994). The synchrotron production of photons with
radiative corrections. {\it Phys. Lett. A} {\bf 189}, 227\\

Pardy, M. (1995). The finite-temperature gravitational \v{C}erenkov
radiation. {\it International Journal of Theoretical Physics} {\bf 34}(6),
951.\\

Pardy, M. (1997). \v Cerenkov effect and the Lorentz Contraction.
{\it Phys. Rev. A} {\bf 55}, No. 3 , 1647-1652.\\

Pardy, M. (2000). Synchrotron production of photons by a two-body
system. {\it International Journal  of Theoretical Physics} {\bf 39},
No. 4, 1109; hep-ph/0008257. \\

Pardy, M. (2002). Largest terrestrial electromagnetic pulsar.
{\it International Journal of Theoretical Physics} {\bf 41}(6),
1155.\\

Rohlf, J. W. (1994). {\it Modern Physics from $\alpha$ to $Z^{0}$}
John Wiley \& Sons, Inc. New York. \\

Sokolov, A. A., Ternov, I. M.,  \v Zukovskii, V. \v C and Borisov, A. V.
(1983). {\it Quantum Electrodynamics}, Moscow University Press.
(in Russian).\\

Schwinger, J., Tsai W.Y. and Erber, T. (1976).
Classical and quantum theory of synergic Synchrotron-\v{C}erenkov
radiation. {\it  Annals of  Physics} ({\it New York}) {\bf 96},
303.\\

Schwinger, J. (1970). {\it Particles, Sources and Fields}, Vol. I,
Addison-Wesley, Reading, Mass..\\

Schwinger, J., (1973). {\it Particles, Sources and Fields}, Vol. II,\\
Addison-Wesley, Reading, Mass..\\

Schwinger, J. (1949). On the classical radiation of accelerated electrons.
{\it Phys. Rev}. {\bf 75}, No. 12, 1912 - 1925.\\

Schwinger, J. (1945). On radiation by electrons in a betatron.
{\it preprint LBNL-39088, CBP Note-179, UC-414}.\\

Tamm, I. E. and Frank I. M. (1937). The coherent radiation of fast electron
in a medium. {\it Dokl. Akad. Nauk USSR}  {\bf 14}
(1937), 107.\\

Van de Graaff, R. J. (1931). A 1,500 000 volt electrostatic generator.
{\it Phys. Rev.} {\bf 38}, 1919 - 20. \\

Wille, K. (2000). {\it The Physics of Particle Accelerators - an
introduction},
First ed. in English, Oxford University Press. \\

Winick, H. (1987). Synchrotron radiation. {\it Scientific American} {\bf 257},
November, pp. 72 -81.

\end{document}